\documentclass[traditabstract]{aa} 
\usepackage{graphicx}
\usepackage{txfonts}
\usepackage{url}

\newcommand{\teff}{\mbox{$T_{\rm eff}$}}
\newcommand{\logg}{\mbox{$\log g$}}

\def\ms{\,m\,s$^{-1}$}         
\def\Rsol{$R_\odot$}             
\def\teff{$T_{\rm eff}$}
\def\logg{$\log g$}

\begin{document}

\title{$H$-band thermal emission from the 19-hour period planet WASP-19b
\thanks{Based on data collected with the VLT/HAWKI instrument at ESO 
Paranal Observatory, Chile (programs 083.C-0377(A)).}\fnmsep\thanks{The 
photometric time-series used in this work are only available in electronic form 
at the CDS via anonymous ftp to  cdsarc.u-strasbg.fr (130.79.128.5) or via 
http://cdsweb.u-strasbg.fr/cgi-bin/qcat?J/A+A/}}

\titlerunning{}

\author{D.~R.~Anderson
        \inst{1}
	\and
        M.~Gillon
	\inst{2,3}
	\and
	P.~F.~L.~Maxted
        \inst{1}
	\and
	T.~S.~Barman
	\inst{4}
	\and
	A.~Collier Cameron
	\inst{5}
	\and
	C.~Hellier
	\inst{1}
	\and
	D.~Queloz
	\inst{3}
	\and
	B.~Smalley
	\inst{1}
	\and
	A.~H.~M.~J.~Triaud
	\inst{3}
       }

\institute{Astrophysics Group, Keele University, Staffordshire ST5 5BG, UK\\
           \email{dra@astro.keele.ac.uk}
           \and
           Institut d'Astrophysique et de G\'eophysique,  Universit\'e de 
	     Li\`ege,  All\'ee du 6 Ao\^ut 17,  Bat.  B5C, 4000 Li\`ege, Belgium
           \and
           Observatoire de Gen\`eve, Universit\'e de Gen\`eve, 51 Chemin des 
             Maillettes, 1290 Sauverny, Switzerland
           \and
           Lowell Observatory, 1400 West Mars Hill Road, Flagstaff, 
	     AZ 86001, USA
           \and
           School of Physics and Astronomy, University of St. Andrews, 
             North Haugh, Fife KY16 9SS, UK}

\date{Received February 9, 2010; accepted March 10, 2010}
\authorrunning{D. R. Anderson et al.}
\titlerunning{Thermal emission from exoplanet WASP-19b}

 
\abstract
{We present the first ground-based detection of thermal emission from 
an exoplanet in the $H$-band. Using HAWK-I on the VLT, we observed an 
occultation of WASP-19b by its G8V-type host star. 
WASP-19b is a Jupiter-mass planet with an orbital period of only 19 h, and 
thus, being highly irradiated, is expected to be hot. 
We measure an $H$-band occultation depth of $0.259^{+0.046}_{-0.044}$\,\%, which
corresponds to an $H$-band brightness temperature of $T_{H} = 2580 \pm 125$\,K. 
A cloud-free model of the planet's atmosphere, with no redistribution of energy 
from day-side to night-side, under predicts the planet/star flux density ratio 
by a factor of two. 
As the stellar parameters, and thus the level of planetary irradiation, 
are well-constrained by measurement, 
it is likely that our model of the planet's atmosphere is too simple. }


\maketitle

%

\section{Introduction}
For a planet that is occulted by its host star, we can measure the planet's 
emergent flux and thus determine its brightness temperature without spatially 
resolving the system (e.g., Charbonneau et al. 2005; Deming et al. 2005). 
From this, we can determine a planet's albedo and the efficiency with which 
energy is redistributed from the day-side to the night-side of the planet 
(e.g., Barman et al. 2008; Alonso et al. 2009a).
By measuring occultations over a range of wavelengths, we can map the planet's 
spectral energy distribution (SED) and can thus  
infer its chemical composition (e.g., Swain et al. 2009a), 
investigate whether a temperature inversion exists 
(i.e., a stratosphere; e.g., Charbonneau et al. 2008; Knutson et al. 2008), 
and constrain the albedo and the day/night energy redistribution 
more strongly.

The {\it Spitzer Space Telescope} (Werner et al. 2004) has measured planetary 
thermal emission in the 3.6--24\,$\mu$m range for 15 exoplanets
(e.g., Deming et al. 2009; Cowan \& Agol 2010; and references therein).
These observations appeared to identify two classes of short-period, giant 
planets, those with stratospheres and those without, depending on whether
there are upper-atmosphere absorbers, and hence 
the planet's effective temperature (e.g., Fortney et al. 2008). 
Retarded cooling produced by opaque atmospheres is one possible cause of 
the anomalously large measured radii (compared to the predictions of simple 
models) of several exoplanets (e.g., Burrows et al. 2007).

For some inflated planets, such as WASP-17b (Anderson et al. 2010), tidal 
heating from the circularisation of an eccentric orbit seems a more likely 
explanation. 
Measuring orbital eccentricity is essential to constrain the current and past 
rates of tidal heating, and the timing of occultations is essential for 
constraining the eccentricity.

The {\it Spitzer} measurements probe only the uppermost layers of the exoplanet 
atmospheres studied to date because their SEDs peak in the 2--4\,$\mu$m 
range. 
Observations at shorter wavelengths are needed to probe the deeper layers of the 
atmosphere where energy redistribution takes place (Burrows et al. 2007), and 
to provide greater leverage when trying to estimate the slope of the temperature 
structure.
Giant planets with orbital periods of a day and shorter are starting to be 
discovered (Hebb et al. 2009; Hellier et al. 2009; Hebb et al. 2010).
With SED peaks around 1\,$\mu$m, measurements bluewards of {\it Spitzer} 
wavelengths are even more important for these planets to estimate  
the bolometric luminosity and thus measure the temperature, albedo, and energy 
redistribution efficiency. 

{\it Spitzer's} wavelength coverage is invaluably extended by ground-based 
facilities, which have measured occultations in the near-infrared (near-IR) 
detect: 
OGLE-TR-56b ($z'$-band, Sing \& L\'opez-Morales 2009), 
TrES-3b ($K$-band, de Mooij \& Snellen 2009), 
and CoRoT-1b (2.09\,$\mu$m, Gillon et al. 2009a; and $K_{s}$-band, Rogers et al. 
2009).
The {\it Hubble Space Telescope} has also measured occultations in the 
near-IR for the two exoplanets with the brightest host stars 
(HD\,189733b, Swain et al. 2009a; and HD\,209458b, Swain et al. 2009b).
Both, {\it CoRoT} (Alonso et al. 2009a, 2009b; Snellen et al. 2009) 
and {\it Kepler} (Borucki et al. 2009) 
are now also measuring occultations from space in the optical. 

In this paper, we present the first ground-based detection of $H$-band thermal 
emission from an exoplanet. Our target, WASP-19b (Hebb et al. 2010), is a 
Jupiter-mass planet in a 19-h orbit around a G8V-type star, and is 
the shortest-period exoplanet currently known.


\section{Observations}
We observed an occultation of WASP-19b by its $H_{\rm mag}=10.6$ host star with 
the cryogenic, near-IR imager HAWK-I 
(Pirard et al. 2004, Casali et al. 2006), mounted on {\it Yepun} of the Very 
Large Telescope (Paranal, Chile). 
HAWK-I is composed of four Hawaii-2RG chips, each measuring 2048x2048 pixels. 
Its pixel scale is 0.106"/pixel, 
providing a total field of view of 7\farcm5 x 7\farcm5. Observations were 
obtained on 2009 May 03 from  23h06 to 05h17 UT. The broadband $H$-band filter 
was used ($\lambda_{\rm eff}$ =  1.620\,$\mu$m, FWHM = 0.289\,$\mu$m). 
The airmass ranged from 1.09 to 2.62 during the run and the transparency 
conditions were good. A total of 405 exposures, each comprising 10 
integrations of 1.26\,s, were obtained. 
We used a pattern of six offsets, with the aim of producing an accurate sky map 
for each image from the neighbouring images.
For a given offset, we kept the stars on the same pixels to minimise the effect 
of the small-scale spatial variations in the sensitivity of the detector chips.
As the pointing was changed once early in the run, each star sampled a total of 
12 detector positions. To avoid saturation of the target and reference stars, 
the telescope was heavily defocused, resulting in asymmetric stellar images with
FWHMs of 13--28\arcsec. 

In our analysis, we used only the images obtained with the $Q1$ chip, which
contained WASP-19 and several reference stars. After a standard pre-reduction 
(dark subtraction and flat-field division), a localised smoothing was applied to
each image: the count level of each pixel was compared to the median count level
of the neighbouring pixels and, if the difference exceeded a threshold of 
4\,$\sigma$ (or 30\,$\sigma$ for pixels belonging to stellar images), then the 
pixel's count level was set to the median of its neighbours. 
At this stage, a sky map was constructed for and removed from each image using a
median-filtered set of the adjacent images taken at different offsets. 
For each of the 12 offsets, aperture photometry was then performed using the 
{\tt IRAF/DAOPHOT}\footnote{{\tt IRAF} is distributed by the National Optical 
Astronomy Observatory, which is operated by the Association of Universities for 
Research in Astronomy, Inc., under cooperative agreement with the National 
Science Foundation.} software (Stetson, 1987). 
An aperture radius of 30 pixels was used. A similar flux extraction was also 
performed on the non-sky-subtracted images and found to provide a more reliable 
result. 
We attribute this to the incomplete removal of the large, asymmetric stellar 
images from the sky maps.
We thus decided to use the fluxes extracted from the non-sky-subtracted images 
in our analysis, though we did measure the sky level in an annulus and 
subtract it from the stellar aperture.

After a careful choice of reference stars, differential photometry was performed
and a light curve was produced for each of the 12 offsets (Fig.~1). 
The light curve of one offset is signifcantly poorer than the 11 others, 
probably because of a detector defect in the image of either 
WASP-19 or a reference star, and we discarded this light curve from
our analysis.
For each of the 11 remaining  light curves, the scatter is much larger during 
one half-hour period. This period corresponds to the 
minimum of the FWHM of the stellar images and to the maximum pixel value (above 
30\,kADU) for the target star. The larger scatter in 
this portion of the light curves is therefore probably caused by a non-linearity 
effect, and so we discarded these data. 
The 276 measurements remaining (from an original 405 measurements) after 
rejection are shown in Fig.~1.

Each light curve varies differently with time, and we found that these 
variations are strongly correlated with the background amplitudes.
We chose not to simply detrend the light curves for external parameters and
analyse the resulting corrected light curves. 
To avoid underestimating the error bars of our final parameters, we
included trend models in our global analysis (Sect. \ref{data-analysis}).

\begin{figure}
\label{fig:a}
\centering                     
\includegraphics[width=9cm]{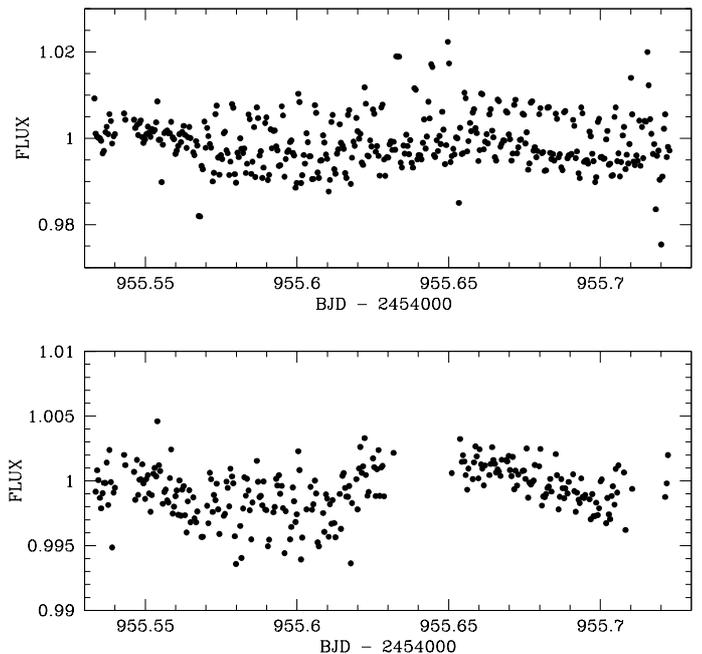}
\caption{ $Top$: The 405 HAWK-I measurements from the twelve pointing offsets. 
The light curve from each offset was given the same normalisation factor. 
$Bottom$: The 276 measurements remaining after the rejection of one offset and 
an half-hour period that exhibited a non-linearity issue.
The light curves were normalised on a per-offset basis.}
\end{figure}

\section{Data analysis}
\label{data-analysis}
To place as many observational constraints as possible on the 
occultation parameters, 
we performed a global analysis of our HAWK-I occultation photometry combined 
with the 34 radial velocities (RVs) and the FTS $z$-band transit light curve
presented in Hebb et al. (2010). In addition, the SuperWASP transit epoch 
reported by Hebb et al. (2010) was used to constrain the orbital period of
the planet. These data were adopted as input of the adaptative Markov-Chain 
Monte Carlo (MCMC) algorithm presented in Gillon et al. (2009a, 2009b). 
This MCMC 
implementation uses the Metropolis-Hasting algorithm (e.g., Carlin \& Louis 
2008) to sample the posterior probability distribution of adjusted parameters 
for a given model.
Our  model was based on an occulting star and a transiting planet on
a Keplerian orbit about their common centre of mass. 
We used a classical Keplerian model for the RVs obtained outside of transit (we 
discarded the single RV obtained during transit). 
To model the eclipse photometry, we used the photometric eclipse model of Mandel 
\& Agol (2002), multiplied by a systematic effect model. 
 
For the FTS transit, a quadratic limb-darkening law was assumed. Quadratic 
coefficients $u_1$ and $u_2$ were interpolated from the tables of Claret (2004)
for the Sloan $z'$-filter and for \teff\ $=5500\pm100$~K, 
\logg~$ = 4.5\pm0.2$, and [Fe/H] = $0.02\pm0.09$ (Hebb et al. 2010). 
We obtained $u_1 = 0.250 \pm 0.015$ and $u_2 = 0.304 \pm 0.008$. We allowed the 
quadratic coefficients $u_1$ and $u_2$ to float in our MCMC analysis, using as 
jump parameters\footnote{Jump parameters are the model parameters that are 
randomly perturbed at each step of the MCMC.} 
the combinations $c_1 = 2 \times u_1 + u_2$  and $c_2 = u_1 - 2 \times u_2$ 
to minimise the correlation of the uncertainties (Holman et al. 2006). 
To obtain a limb-darkening solution consistent with theory, we added to our 
merit function the Bayesian penalty 
  \begin{equation}
  BP_{\rm \textrm{ }  ld} = \sum_{i=1,2} \bigg(\frac{c_i - c'_i}{\sigma_{c'_i}} 
  \bigg)^2\textrm{,}
  \end{equation}
where $c'_i$ is the initial value deduced for the coefficient $c_i$ 
and $\sigma_{c'_i}$ is its error.

For each light curve, the eclipse model was multiplied by a trend model to take 
into account known low-frequency noise sources (instrumental and stellar). For 
the FTS transit light curve, we modelled this possible trend as a quadratic 
time-dependent polynomial. Given the small number of points in the HAWK-I 
light curves obtained prior to the re-pointing of the telescope, 
we used a simple 
linear time-function as a trend model for them. For the other five HAWK-I light 
curves, our adopted trend model was a combination of a quadratic function of 
time and a quadratic function of the background amplitude. 
As explained in Gillon et al. (2009b), the coefficients of the trend models are 
not jump parameters of the MCMC, but are rather determined by least squares 
minimisation at each step of the MCMC. 
We followed the procedure described in Winn et al. (2008) to check for 
correlated noise in the photometric time-series and to scale the error bars 
accordingly. 
For the RVs, the 
quadratic addition of a jitter noise of 13 m s$^{-1}$ was needed to obtain a 
residual $rms$ in good agreement with the mean error in the measurements.  

The jump parameters in our MCMC simulation were: 
the planet/star area ratio $(R_p /R_s )^2$; the $H$-band occultation depth; 
the transit width (from first to last contact) $T_{14}$; the modified impact 
parameter $b' = a \cos{i}/R_\ast$ (Gillon et al. 2009a); 
the orbital period $P$; the time of mid-transit $T_0$; 
the two Lagrangian parameters $e \cos{\omega}$ and $e \sin{\omega}$, 
where $e$ is the orbital eccentricity and $\omega$ is the argument of 
periastron; and $K_2=K \sqrt{1-e^2/}P^{1/3}$ (Gillon et al. 2009a), 
where $K$ is the semi-amplitude of the radial stellar reflex velocity. 
We assumed a uniform prior distribution for all of these jump parameters. 
The merit function used in our analysis was the sum of both the $\chi^2$ of  
each time-series and the Bayesian penalty presented in Eq.~1.

At each step of the MCMC, stellar mass was determined using the stellar mass 
calibration relation of Torres et al. (2009), which was derived from 
detailed studies of detached stellar binaries. The parameters in this relation 
are \logg, \teff\, and [Fe/H], though, following the method of Enoch et al. 
(2010), we recast it in terms of stellar density, $\rho_{*}$, instead of \logg.
Stellar density is constrained by observation, as it depends on the shape of the
transit light curve and the eccentricity of the orbit, which is constrained by
the RVs and occultation timing.
The values of \teff\ and [Fe/H] were drawn from the distributions 
$N(5500,100^2)$\,K and $N(0.02,0.09^2)$, respectively, 
where $N(\mu,\sigma^2)$ is a normal distribution with mean $\mu$ and variance 
$\sigma^2$.
To account for the uncertainty in the parameters of the stellar 
calibration law, the values of these parameters were randomly drawn at each
step of the MCMC from the normal distributions presented in Torres et 
al. (2009).

We first perfomed a single MCMC to assess the levels of correlated noise 
in the photometry and jitter noise in the RVs, and to scale the measurement 
error bars accordingly. 
This MCMC consisted of $10^5$ steps, with the first 20\,\% considered as its
`burn-in' phase and discarded. 
Five new MCMCs (10$^5$ steps each, with 20\,\% burn-in) were then performed 
using the scaled error bars. 
The good convergence and mixing of these five MCMCs was verified using 
the Gelman and Rubin (1992) statistic ($R$ parameter $<$ 1\,\%). Finally, the 
inferred value and error bars of each parameter were obtained from its 
marginalised posterior distribution. Table 1 shows the deduced values of the 
jump and derived parameters.
The best-fit planet model is shown in Figs.~2 and 3. As can be seen in
Fig.~3 and Table 1, the occultation is clearly detected, with a depth of 
$0.259^{+0.046}_{-0.044}$\,\%.
The $rms$ of the HAWK-I residuals is 0.12\,\%, and 0.047\,\% after binning 
in intervals of 10 measurements. 

To investigate the robustness of the marginal detection of a non-zero 
eccentricity ($e = 0.016^{+0.015}_{-0.007}$), we also perfomed an MCMC with 
eccentricity fixed to zero. 
To compare the eccentric and circular models, we computed the Bayes factor
(e.g., Carlin \& Louis 2008), which is the ratio of the marginal likelihoods 
(e.g., Chib \& Jeliaskov 2001) of the two models.
The deduced Bayes factor of 88 indicates that the data are more consistent with
the eccentric model, though not decisively so.

\begin{table}
\label{tab:params}
\begin{tabular}{lc}
\hline
Parameter (unit)  & Value	\\ \noalign {\smallskip}
\hline \noalign {\smallskip}

\multicolumn{2}{l}{{\it Fitted model parameters (jump parameters)}}\\
\noalign {\smallskip}
\hline \noalign {\smallskip}
$R_p^2/R_s^2$		& $0.01982^{+ 0.00050}_{- 0.00049}$	\\
\noalign {\smallskip}
$b'=a\cos{i}/R_\ast$	& $ 0.631^{+ 0.026}_{- 0.029} $		\\
\noalign {\smallskip}
$T_{\rm 14}$ (days)	& $ 0.06506^{+0.00072}_{-0.00071}$	\\
\noalign {\smallskip}
$ T_0$ (HJD)		& $ 2454776.91490 \pm 0.00019$		\\
\noalign {\smallskip}
$ P$ (days)		& $ 0.7888393 \pm 0.0000045$		\\
\noalign {\smallskip}
$K_2$ (m\,s$^{-1}$ days$^{1/3}$)	& $236.9^{+5.5}_{-5.6}$	\\
\noalign {\smallskip}
$e\cos{\omega}$		& $0.0069^{+0.0024}_{-0.0027}$		\\
\noalign {\smallskip}
$e\sin{\omega}$		& $0.005 \pm 0.021$			\\
\noalign {\smallskip}
$H$-band $F_p/F_\ast$	& $0.00259^{+0.00046}_{-0.00044} $	\\
\noalign {\smallskip}
$c_{1}$$^{\dagger}$		& $ 0.810 \pm 0.030$  			\\
\noalign {\smallskip}
$c_{2}$$^{\dagger}$		& $ -0.357 \pm 0.022$ 			\\
\noalign {\smallskip}

\hline \noalign {\smallskip}

\multicolumn{2}{l}{{\it Dependant parameters deduced from the above}}\\
\noalign {\smallskip}
\hline \noalign {\smallskip}
$K$ (\ms)			& $256.5  \pm 6.0$		\\
\noalign {\smallskip}
$b_{\rm transit}$	& $ 0.628^{+ 0.025}_{- 0.029}$		\\
\noalign {\smallskip}
$b_{\rm occultation}$	& $ 0.633^{+ 0.033}_{- 0.034}$		\\
\noalign {\smallskip}
$T_{\rm occultation}$ (HJD) & $ 2454777.31281^{+0.00013}_{-0.00014}$\\
\noalign {\smallskip}
$a$ (AU)		& $0.01659^{+ 0.00040}_{- 0.00039}$	\\
\noalign {\smallskip}
$i$ (\degr)		& $ 80.07^{+0.72}_{-0.70}$		\\
\noalign {\smallskip}
$e$			& $ 0.016^{+0.015}_{-0.007}$		\\
\noalign {\smallskip}
$\omega$ (\degr)	& $33^{+44}_{-101}$			\\
\noalign {\smallskip}
$M_\ast$ ($M_\odot$)	& $ 0.977^{+0.073}_{-0.067}$ 		\\
\noalign {\smallskip}
$ R_\ast$ ($R_\odot$)	& $0.975^{+ 0.038}_{- 0.037}$		\\
\noalign {\smallskip}
$\rho_* $ ($\rho_\odot$)& $1.054^{+ 0.095}_{- 0.085}$		\\
\noalign {\smallskip}
$u_1$			& $0.253^{+ 0.015}_{- 0.014}$ 		\\
\noalign {\smallskip}
$u_2$			& $0.3049 \pm 0.0083$			\\
\noalign {\smallskip}
$M_p$ ($M_J$)		& $1.166^{+ 0.064}_{- 0.061}$		\\
\noalign {\smallskip}
$R_p$ ($R_J$)		& $1.338 ^{+ 0.060}_{- 0.059}$ 		\\
\noalign {\smallskip}
$\rho_p$ ($\rho_{J}$)	& $0.486^{+0.059}_{+0.051}$ 		\\
\noalign {\smallskip}
\hline\\ \noalign {\smallskip}
\end{tabular}
\caption{WASP-19 system parameters and 1-$\sigma$ error limits. 
$^\dagger$Constrained parameter; see text for details.}
\end{table}

\begin{figure}
\label{fig:b}
\centering                     
\includegraphics[width=9cm]{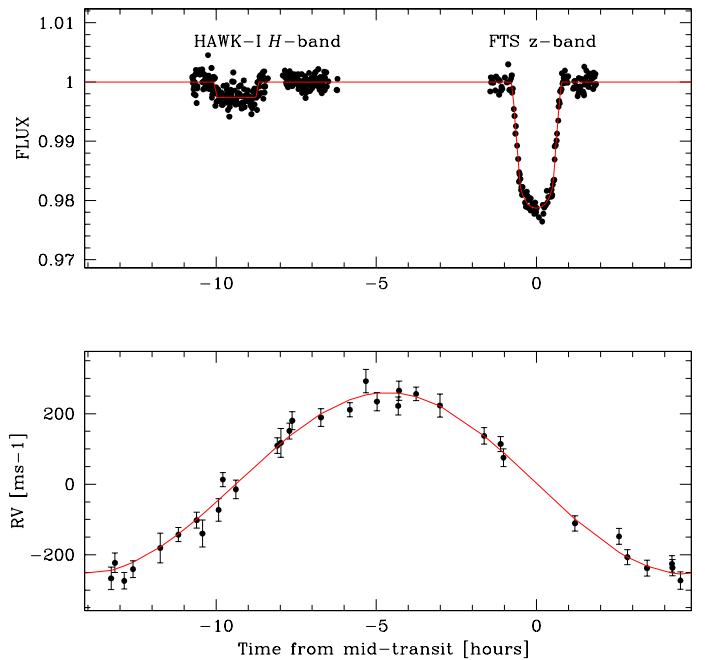}
\caption{$Top$: Photometry of the planetary transit (FTS data, from Hebb et al. 
2010) and the planetary occultation (HAWK-I data), folded on the transit 
ephemeris. The best-fit light curve model is superimposed. 
$Bottom$: Radial velocities (from Hebb et al. 2010) folded on the best-fit 
transit ephemeris. The best-fit RV model is superimposed.}
\end{figure}

\begin{figure}
\label{fig:c}
\centering                     
\includegraphics[width=9cm]{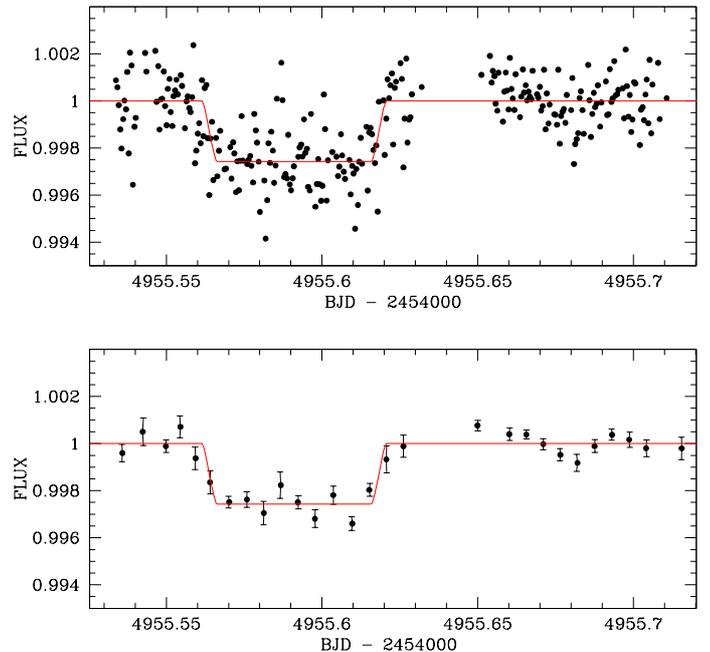}
\caption{HAWK-I photometry corrected by the best-fit trend model for each 
time-series. The best-fit occultation model is superimposed. 
$Top$: unbinned. $Bottom$: binned every 10 measurements. 
Our global solution indicates a mid-occultation time of BJD=2\,454\,955.5905. 
If a circular orbit is assumed, the transit ephemeris of Table~1  predicts a 
time of BJD=2\,454\,955.5870, earlier by 0.0035\,d 
(=5\,min =0.35$\times$\,a minor {\it tic} increment).}
\end{figure}

\section{Discussion}
We have measured a decrease in the $H$-band flux from the 
WASP-19 system of $0.259^{+0.046}_{-0.044}$\,\%, which we attribute to the 
occultation of day-side planetary thermal emission by the host star.

Our global solution suggests that the planet's orbit may be non-circular, the 
value of $e\cos{\omega}$ being non-zero at the 2.6-$\sigma$ level; 
the constraint on $e\sin{\omega}$ is far weaker. 
The measurement of additional occultations and RVs will constrain 
$e\cos{\omega}$ and $e\sin{\omega}$ more strongly. 
A non-circular orbit may indicate that tidal heating is responsible for 
WASP-19b's inflated radius (e.g., Ibgui et al. 2009).

The measured occultation depth corresponds to an $H$-band brightness 
temperature of $T_{H} = 2580 \pm 125$\,K. 
To calculate $T_{H}$, we defined the product of the planet/star area ratio and 
the ratio of the bandpass-integrated planetary to 
stellar surface photon fluxes, corrected for transmission\footnote{The 
transmission of the atmosphere, telescope, instrument, and detector were taken 
into account. \url{http://www.eso.org/observing/etc/}}, to be 
equal to the measured occultation depth (e.g., Charbonneau et al. 2005). 
We assumed the planet to emit as a black body and, for the 
star, we used a model spectrum of a G5V star (Pickles 1998), normalised to 
reproduce the integrated flux of a black body with \teff~$ = 5500$\,K 
(Hebb et al. 2010). 
The uncertainty in $T_{H}$ only takes into account the uncertainty in the 
measured occultation depth.

In Fig.~4, the measured $H$-band flux density ratio is compared to a model 
atmosphere spectrum of the planet (Barman et al. 2005), which was based on
parameter values from Table~1. 
The model is cloud-free and adopts the nominal day-side scenario, in 
which the incident stellar energy remains entirely on the day-side. 
Given the planet's small orbital distance from the G8V-type host star, the 
irradiation is quite strong ($T_{\rm eq}\approx2400$\,K, day-side only) and, 
in these models, it is common to see a nearly isothermal photosphere with a 
large temperature inversion at low pressures.  In the model explored here, the 
inversion layers have temperatures exceeding 3000\,K. 
Despite these high temperatures, the inversion does not extend deep enough into 
the photosphere to match the observed $H$-band flux; the model under predicts 
the flux by roughly a factor of two.

To reconcile our model's prediction with the occultation measurement, 
one would need to increase the luminosity of the star by a factor of two, 
for example, by increasing the stellar temperature by 1000\,K or increasing the 
host star radius by 0.4\,\Rsol\ (or some combination of these). 
These values of the stellar parameters are excluded by observations 
(Hebb et al. 2010; this work). 
The most probable explanation of the poor match to the planet flux is therefore 
related to the model atmosphere.
 
It is possible that the day-side atmosphere is much closer to pure 
radiative-convective equilibrium and experiences little day-side redistribution. 
This zero-redistribution scenario has been shown to result in a hotter 
substellar point and deeper occultations than the uniform day-side-only 
model (Barman et al. 2005). 
As more data are obtained, in particular at other 
near-IR wavelengths, we will explore a wide 
variety of atmospheric phenomena (e.g., clouds, depth-dependent energy 
redistribution, and photochemistry). 

Swain et al. (2010) found a strong, unexpected 3.25-$\mu$m emission feature in 
the day-side spectrum of HD\,189733b, which they attributed to 
non-local-thermodynamic-equilibrium (non-LTE) CH$_4$ emission.
Future work should explore whether non-LTE chemistry is responsible for the 
unexpectedly high $H$-band flux of WASP-19b. 
For now, our measured $H$-band flux presents an interesting puzzle for 
irradiated atmospheric models.

\begin{figure}
\label{fig:d}
\centering                     
\includegraphics[width=9cm]{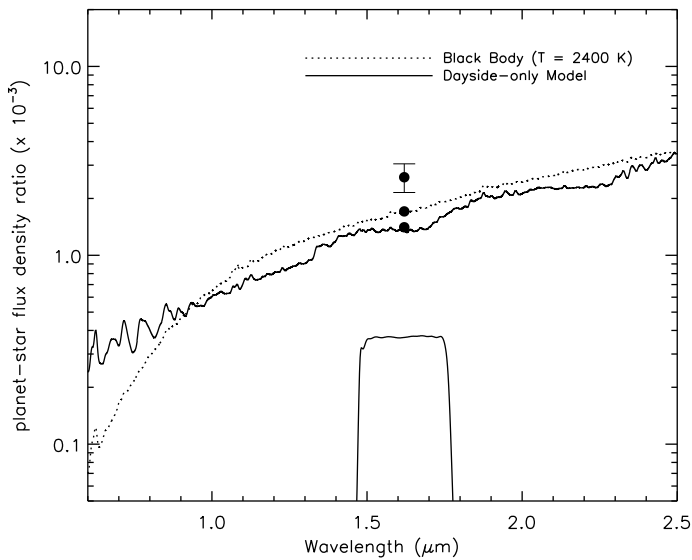}
\caption{Synthetic planet/star flux ratios of a cloud-free model that assumes 
solar abundances and no redistribution of incident stellar flux to the 
night-side (solid line).  The effective temperature of the star was taken to be
\teff\ = 5500K (Hebb et al. 2010); all other parameters are taken from 
Table~1. The flux 
ratio for a black-body planet (at 2400K) is indicated by a dotted line. Solid 
symbols are the $H$-band integrated flux ratios; symbols without error bars 
correspond to the model points.  The HAWK-I $H$-band filter response curve is 
also shown.}
\end{figure}

\begin{acknowledgements} 
M. Gillon acknowledges support from the Belgian Science Policy Office in the 
form of a Return Grant. 
\end{acknowledgements}

\end{document}